\documentclass[aps,graphicx,reprint,amsmath,amssymb,groupedaddress]{revtex4-1}
\usepackage{graphicx}
\usepackage{bbm}
\usepackage{amsmath}
\usepackage{mathrsfs}
\usepackage{amsfonts}
\usepackage{graphicx}
\usepackage{amscd}
\usepackage{color}
\begin{document}
\title{The periodic table of topological insulators and superconductors: the loop sequence approach}
\author{Chunbo Zhao}\email[]{cbzhao@semi.ac.cn}
\address{State Key Laboratory of Superlattices and Microstructures, Institute of Semiconductors, Chinese Academy of Sciences, P. O. Box 912, Beijing 100083, People's Republic of China}

%\date{\today}

\begin{abstract}
In this work, a complete homotopic interpretation on the periodic table of topological insulators and superconductors has been derived by establishing the loop sequence of the corresponding classifying spaces. In our approach, each classifying spaces has been regarded as a fiber bundle defined on one point, following that the loop relationship between the classifying spaces in the same symmetric class has been built and proved. Meanwhile, the loop sequence between the classifying spaces with the same spatial dimension has also been developed based on the minimal geodesic approximation. Finally, by building the loop sequence for general dimensions and symmetric classes, a complete homotopic understanding about the whole periodic table of any classifying spaces is achieved. Our approach provides a unified way to interpret the topological relationship between any free-fermion systems.
\end{abstract}
\date{\today}
\maketitle
\section{Introduction}
Recently, topological insulators and superconductors have drawn a great attention in the field of condensed matter physics\cite{hasan2010colloquium,moore2010birth,qi2010quantum,qi2011topological}. They are gapped phases of fermions with topologically protected boundary modes. These boundary states, however, are gapless and protected against arbitrary perturbations. They can be described by the topological band theories if excluding the interactions. The classification of gapped free fermion system has been systematically developed by means of K-theory\cite{kitaev2009periodic,stone2011symmetries,wen2012symmetry}, minimal Dirac representation\cite{ryu2010topological}, or nonlinear $\sigma$ model\cite{schnyder2008classification} approaches, respectively. But for the interacting systems, it is generally difficult to directly study the topological classification of gapped interacting Hamiltonians. Qi $et al$ \cite{qi2008topological} obtained the classification based on the topological response field theories, i.e., Chern-Simons field theory under the constraint of discrete symmetries, in which the topological order parameters were included as a coefficient and can be expressed with Green functions\cite{volovik2009universe,wang2010topological,gurarie2011single}. However, the classification is usually very difficult for practical evaluation since the topological order parameters are associated with the integral of Green function within the entire frequency domain. Recently, Wang $et al$\cite{wang2012simplified}  greatly simplified the integral topological invariants by showing that the essential topological information can be extracted just from the zero-frequency Green function or topological Hamiltonians\cite{wang2013topological}, which is in analogy with the free fermion systems. Therefore, the topological classification of the interacting systems can be directly studied with the topological Hamiltonians as developed by Wang. Thus it is a prerequisite to fully understand the topological structure of the periodic table for the non-interacting topological insulators and superconductors , in order to furthermore study the classification of corresponding interacting ones  such as topological Mott and Anderson insulators\cite {raghu2008topological,li2009topological}.
\par
 The periodic table of non-trivial free-fermion topological states is presented in Table \ref{table1}, in which the ten symmetric classes has been grouped into two classes called the $complex$ and $real$ cases, respectively\cite{ryu2010topological}. The classification of the Hamiltonian is based on the symmetry principle, and is first developed with random matrix theory by Altland and Zirnbauer\cite{altland1997nonstandard,zirnbauer1998riemannian} in physics community.
  It is known from Table \ref{table1} that there exist precisely five distinct classes of topological states in every spatial dimension.  And it is easily to obtain clear 2-fold and 8-fold pattern with the variation of spatial dimension $d$ (the same row of Table.\ref{table1}) for the $complex$ and $real$ cases, respectively. Moreover, by looking into the shift along the symmetric axis $s$ (the same column of Table \ref{table1}), one could also get similar 2-fold and 8-fold period pattern for  the $complex$ and $real$ cases, respectively. The beautiful periodic pattern illustrated above can be understood based on the abstract real K-theory\cite{kitaev2009periodic} with Bott periodic theorem\cite{bott1959stable}. However, a relatively simple interpretation will be beneficial for physicists to easily catch up the fundamental basis for the topological classification.  For this purpose, M. Stone, $et al$ \cite{stone2011symmetries} has given a relatively simple homotopy interpretation about the topological relationship between the elements along axis $s$ for $d=0$ case and the periodic pattern was therefore obtained. But there is no a simple homotopic explanation for the spatial dimension $d$ dependence yet so far in the same row of the table.
 \begin{table}
\caption{\textbf{Periodic table for the classification of topological insulators and superconductors. The second and third rows are associated with two distinct super-families in the Altland-Zirnbauer classification, a set of two is associated with the unitary group and a set of eight is associated with the real orthogonal group. The first column is the Cartan label of classifying spaces, with chiral classes denoted by bold letters. The symmetry classes can be represented by an integer $s$ defined modulo 2 and 8 for the complex and real cases, respectively. The symbols $\mathbb{Z}$ and $\mathbb{Z}_2$ indicate that the topologically distinct phases within a given symmetry class of topological insulators (superconductors) are characterized by an integer invariant ($\mathbb{Z}$) or ($\mathbb{Z}_2$) quantity, respectively.}}\label{table1}
\begin{center}
\begin{tabular}{c|c|cccccccc}
  \hline\hline
  AZ&s$\setminus$d & 0 & 1 & 2 & 3 & 4 & 5 & 6 & 7 \\\hline

  A &0 & $\mathbb{Z}$ & 0 & $\mathbb{Z}$ & 0 & $\mathbb{Z}$ & 0 & $\mathbb{Z}$ & 0 \\
  \textbf{AIII}&1& 0 & $\mathbb{Z}$ & 0 & $\mathbb{Z}$ & 0 & $\mathbb{Z}$ & 0 &$\mathbb{Z} $\\\hline

  AI&0 & $\mathbb{Z} $& 0 & 0 & 0 & $\mathbb{Z}$ & 0 & $\mathbb{Z}_2$ &$\mathbb{Z}_2$\\
 \textbf{ BDI}&1& $\mathbb{Z}_2$&$\mathbb{Z}$ & 0 & 0 & 0 & $\mathbb{Z} $& 0 &$ \mathbb{Z}_2$\\
  D &2& $ \mathbb{Z}_2$&$\mathbb{Z}_2$&$\mathbb{Z}$ & 0 & 0 & 0 & $\mathbb{Z}$ & 0\\
  \textbf{DIII}&3 & 0&$\mathbb{Z}_2$&$\mathbb{Z}_2$&$\mathbb{Z}$ & 0 & 0 & 0 & $\mathbb{Z}$\\
  AII &4&  $\mathbb{Z}$& 0&$\mathbb{Z}_2$&$\mathbb{Z}_2$&$\mathbb{Z}$ & 0 & 0 & 0\\
 \textbf{ CII}&5 &0 &  $\mathbb{Z}$& 0&$\mathbb{Z}_2$&$\mathbb{Z}_2$&$\mathbb{Z} $& 0 & 0 \\
  C &6& 0 &0 &  $\mathbb{Z}$& 0&$\mathbb{Z}_2$&$\mathbb{Z}_2$&$\mathbb{Z}$& 0 \\
  \textbf{CI}&7 & 0 &0 &0 &  $\mathbb{Z}$& 0&$\mathbb{Z}_2$&$\mathbb{Z}_2$&$\mathbb{Z} $\\
  \hline\hline
\end{tabular}
\end{center}
\end{table}
 \par
\begin{figure}[h]
  \centering
 \scalebox{0.4} [0.4]{\includegraphics{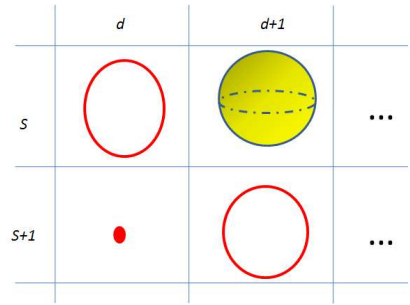}}\\
  \caption{ The schematic loop sequence of the periodic table \ref{table1}. The circle in the second row stands for the classifying space manifold $\mathcal {R}_{s,d}$, which is the loop space of $\mathcal {R}_{s,d+1}$ denoted with a sphere. Meanwhile, it is also the suspension space of $\mathcal{R}_{s+1,d}$ denoted as a point. }\label{1}
\end{figure}

 In this paper, we constructed the topological footing of loop sequence explicitly for any classifying spaces with their homotopy groups denoted as the elements in the periodic table. This approach provides a unified way to interpret the topological relationship between any free-fermion systems, and an explicit topological classification of the system is therefore obtained. By building the loop structure for general dimensions and symmetric classes, one can get a complete homotopic explanation about the whole periodic table. And meanwhile, the 'anomalous' dimension dependence of classification\cite{stone2011symmetries} originating from $\textbf{k}\rightarrow-\textbf{k}$ inversion of the Bloch momentum associated with the antiunitary discrete symmetry can be understood simultaneously. Our key results can be written as $\mathcal {R}_{s+1,d}=\Omega(\mathcal {R}_{s,d})$ and $\mathcal {R}_{s,d+1}=\Sigma(\mathcal {R}_{s,d}) $, where $\mathcal {R}_{s,d}$ stands for the classifying space manifold with $s$ (mod 8) symmetric index and $d$ (mod 8) spatial dimension index, $\Omega$ and $\Sigma$ are the loop space and reduced suspension operators\cite{crossley2005essential}, respectively.  Based on these results, the topological relationship of corresponding Hamiltonians can be illustrated intuitively as in  Fig.\ref{1}. It is clear from Fig.\ref{1} that the neighbouring classifying spaces show similar relation as equator-and-sphere  from the view point of topology. And the same graphic presentation in the diagonal of Fig.\ref{1} demonstrates the 'symmetry cancellation' effect induced by increasing $d$,  with $\mathcal {R}_{s+1,d+1}=\Omega\Sigma(\mathcal {R}_{s,d})=\Omega\Omega^{-1}(\mathcal {R}_{s,d})=\mathcal {R}_{s,d}$, where the mathematic fact $\Sigma=\Omega^{-1}$ has been adopted. These topological relationship exhibited in Fig.\ref{1} will finally lead to the characterization and classification of topological distinct states of matters by using $\pi_{0}(\mathcal {R}_{s-d})$, which will be discussed in detail later. In analogy with the $real$ case, the classifying spaces $C_s$ provide similar topological structures as $\mathcal{R}_s$, whose discussions are omitted here for simplicity.

\par
 This paper is organized as follows.  First, the loop sequence is proven in the same symmetric class stimulated by the 'interpolation' method\cite{qi2008topological}. Second, the loop sequence of symmetric spaces in $d=0$ case is developed by minimal geodesic approximation and homotopy discussions\cite{milnor1963morse}. Finally, the loop sequence based topological relationship between any classifying spaces as listed in Table\ref{table1} is presented and summarized.
\par
\section{Loop sequence of $\mathcal {R}_{s,d}$ for the same symmetry class}

\par
We first explain the topological pattern $\mathcal {R}_{s,d+1}=\Omega^{-1}(\mathcal {R}_{s,d}) $ for classifying spaces in a single row of Table\ref{table1}. In K-theory, the Hamiltonian $\mathcal {H}(\textbf{k})$ is usually viewed as a map from $base~ space$ Brillouin zone parameterized by $\textbf{k}$ to $classifying ~space$ $\mathcal {R}_{s}$\cite{teo2010topological},  with the Brillouin zone assumed to have the topology of sphere for convenience. In our approach, however, an entire geometrical object, denoted as $\mathcal {H}_s^d(\textbf{k})\simeq \mathcal {R}_{s,d}$, is treated by the fiber bundle with a fiber $\mathcal {R}_{s,d}$ defined on one point. For the special case $d = 0$, the gapped Hamiltonian $\mathcal {H}(\textbf{k})$ is identical to the usual real symmetric spaces $\mathcal {R}_{s,0}=\mathcal{R}_s$  that have been studied previously in both mathematics \cite{milnor1963morse,bott1959stable} and physics communities\cite{stone2011symmetries,wen2012symmetry}.
Consequently, the topologically distinct gapped states of free-fermions system can be classified by using $\pi_0(\mathcal {R}_{s,d})$.

\par
    To prove the topological loop relationship  between the neighbouring gapped Hamiltonians $\mathcal {H}(\textbf{k})$ of spatial dimension in the same symmetry class $s$, we need to define two spaces, one of which is called path space and the other one is named homotopy map space. The path space $\Omega(\mathcal {H}_s^d(\textbf{k}),p,q)$ is the set of paths  $\gamma(t)$ on the manifold of Hamiltonian $\mathcal {H}_s^d(\textbf{k})$ that starts from point $p=\gamma(0)$ and ends at point $q=\gamma(\pi)$,

   \begin{equation}\label{path space}
    \Omega(\mathcal {H}_s^d(\textbf{k}),p,q)=\{\gamma(t)\in \mathcal {H}_s^d(\textbf{k})|\gamma(0)=p, \gamma(\pi)=q\}.
 \end{equation}
 Geometrically, one specific $\gamma(t)$ is nothing but a curve parameterized by $t$ on the manifold of $\mathcal {H}_s^d(\textbf{k})$. Subsequently, we need to construct a homotopy map $h^{d}_{s}(\textbf{k},\theta)$ between two Hamiltonian's manifold in the space with one dimension  lower such as $\mathcal {H}_s^{d-1}(\textbf{k})_1$ and $\mathcal {H}_s^{d-1}(\textbf{k})_2$, while preserving the corresponding symmetry. Here the map or interpolation $h^{d}_{s}(\textbf{k},\theta)$ is required to be a 'gapped interpolation' as defined in Ref.\cite{qi2008topological}.  Then we can define the other space called homotopoy map space $\Omega_h(h^{d}_{s}, \mathcal {H}_1, \mathcal {H}_2)$, which  satisfies $h^{d}_{s}(\textbf{k},0)=\mathcal {H}_s^{d-1}(\textbf{k})_1=\mathcal {H}_1$ and $h^{d}_{s}(\textbf{k},\pi)=\mathcal {H}_s^{d-1}(\textbf{k})_2=\mathcal {H}_2$,
   \begin{equation}\label{homotopy set}
    \Omega_h(h^{d}_{s}, \mathcal {H}_1, \mathcal {H}_2)=\{h^{d}_{s}|h^{d}_{s}(\textbf{k},0)=\mathcal {H}_1,h^{d}_{s}(\textbf{k},\pi)=\mathcal {H}_2\}.
   \end{equation}

   \par
    Next, we will show that, for each path $\gamma(t)$ in $path~space$ $\Omega(\mathcal {H}_s^d(\textbf{k}),p,q)$, one can find a homotopically equivalent map $h^{d}_{s}(\textbf{k},\theta)$ in $\Omega_h(h^{d}_{s}, \mathcal {H}_1, \mathcal {H}_2)$ or vice versa, if setting $p=\mathcal {H}_1$ and $q=\mathcal {H}_2$, with $p$ and $q$ to be the any points in the submanifold $\mathcal {H}_1$ and $\mathcal {H}_2$, respectively. For this purpose, we establish a map:
\begin{equation}\label{path space and homotopy set}
    F: \Omega(\mathcal {H}^d_s(\textbf{k}),p,q)\rightarrow \Omega_h(h^{d}_{s}, \mathcal {H}_1, \mathcal {H}_2),
\end{equation}
from the path space to homotopy map space. It should be noted that the map $F$ is actually an isomorphism based on the following analysis. On the one hand, once having a path $\gamma(t)$ from point $p$ to $q$ built, one can define $h_s^d(\textbf{k},t)$ with $h_s^d(\textbf{k},0)=\mathcal {H}_1$ and $h_s^d(\textbf{k},\pi)=\mathcal {H}_2$, so that $h_s^d(t)=\gamma(t)$ when the map $h_s^d(\textbf{k},t)$ is restricted to the connected region from point $p$ to $q$.  On the other hand, if $interpolation$ $h_s^d(\textbf{k},\theta)$ is  built, one can choose any point $p$ in the manifold $\mathcal {H}_1$ and the corresponding point $q$ in $\mathcal {H}_2$ through $h_s^d(\textbf{k},\theta)$, then the path can be written  as $\gamma(\theta)=h^{d}_{s}(\theta)$.  Since the topologically disconnected region of manifold $\mathcal {H}_1$ or $\mathcal {H}_2$ can be deformed continuously to the discrete points correspondingly, thus the map $F$ is an isomorphism from the view point of topology. Intuitively, $h^{d}_{s}(\textbf{k},\theta)$ can be regarded as the homotopy map between two $super~surface$ $\mathcal {H}_1$ and $\mathcal {H}_2$ in manifold $\mathcal {H}_s^d(\textbf{k})$ with parameter $\theta$.
 \par
\begin{figure}[h]
  \centering
 \scalebox{0.5} [0.4]{\includegraphics{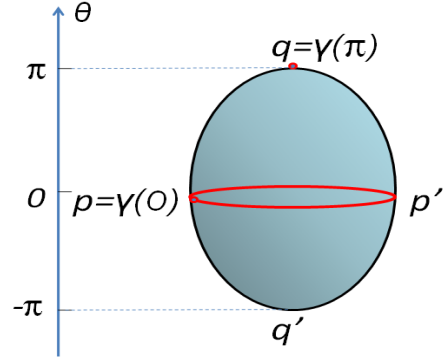}}\\
  \caption{ The schematic sketch of the $interpolation$ between $\mathcal {H}_1$ denoted as the red equator of the sphere and $\mathcal {H}_2$ denoted as the north pole of the sphere (here we assume $\mathcal {H}_2$ a constant Hamiltonian). The curve $\gamma(t)$ from $p$ to $q$ will lead to a closed loop when combining the discrete symmetry partners $p'$ and $q'$, $p\rightarrow q\rightarrow p^{'}\rightarrow q^{'}\rightarrow p$.}\label{shematic}
\end{figure}

  In general, the homotopy map $h_s^d(\textbf{k},\theta)$ for $\theta\in(0,\pi)$ is not a Hamiltonian for the system with one dimension higher in symmetry class $s$ if we replace $\theta$ by a momentum wave vector. Fortunately, it will be one if we combine the discrete symmetric partner for $\theta\in [-\pi,0]$ as seen in Fig.\ref{shematic}, where the south hemisphere is the symmetric partner of the north one. For convenience, we assume that $\mathcal {H}_2$ is a constant referent Hamiltonian that doesn't depend on the momentum $\textbf{k}$, so it can be regarded as one point called north pole. In fact, this process corresponds topologically to the suspension operator $\Sigma$ in mathematics. For example, if the manifold of $\mathcal {H}_s^{d-1}(\textbf{k})_1$ is a circle, then the $H_s^d(\textbf{k})$ manifold for higher dimension will be a sphere, which is the suspension of circle $S^2=\Sigma(S^1)$. Explicitly, we define the $interpolation$ partner for $\theta\in[-\pi,0]$ as $h_s^d(\textbf{k},\theta) = -[C^{\dagger}h_s^d(-\textbf{k},-\theta)C]^{T}$ and $h_s^d(\textbf{k},\theta) =[T^{\dagger}h_s^d(-\textbf{k},-\theta)T]^{T}$ for particle-hole and time-reversal symmetry, respectively. And meanwhile $h_s^d(\textbf{k},\theta)$ is assumed to be gapped for any $\theta\in[-\pi,0]$ with $2\pi$ period in $\theta$. After the above construction, the $interpolation$ for $\theta\in [0,2\pi]$ will become loop spaces as $\gamma(0)=\gamma(2\pi)$, which is parameterized by 'equator' manifold $\gamma(0)=\mathcal {H}_s^{d-1}(\textbf{k})_1$. If we denote loop space as $\Omega_h(h^{d}_{s}, \mathcal {H}_1)$ or $\Omega(\mathcal {H}_s^d(\textbf{k}),p)$ with base point $p=\mathcal {H}_1$,
   then the  isomorphism of $F$ will directly lead to an one-to-one map between homotopy maps and loop space of $\mathcal {H}_s^d(\textbf{k})$ as:
  \begin{equation}\label{loop space}
    M: \Omega_h(h^{d}_{s}, \mathcal {H}_1) \rightarrow \Omega(\mathcal {H}_s^d(\textbf{k}),p).
  \end{equation}
Physically, the adiabatic evolution of $\theta$ from 0 to $2\pi$ defines a cycle of adiabatic pumping in $h_s^d(\textbf{k},\theta)$\cite{qi2008topological}.
  \par
   Since the number of homotopically non-equivalent Hamiltonians $\mathcal {H}_s^{d-1}(\textbf{k})$ in $d-1$ dimension is determined by the homotopically non-equivalent $interpolation$ $h_s^d$,  one can thus classify the $(d-1)$-dimensional Hamiltonians as:
\begin{equation}\label{homotopy of loop of interpolation}
   \pi_0(\mathcal {R}_{s,d-1})=  \pi_0(\Omega_h(h^{d}_{s}, \mathcal {H}_1))=\pi_0(\Omega\mathcal {H}_s^d(\textbf{k}))=\pi_{1}(\mathcal {H}_s^d(\textbf{k}),
\end{equation}
where we have used the standard isomorphism $\pi_n(\Omega X)=\pi_{n+1}(X)$, $X$ is any manifold. In other words, the classifying space $\mathcal {R}_{s,d-1}$ is topologically homeomorphism for the loop space $\Omega(\mathcal {H}_s^d(\textbf{k}),p)$.  Therefore one can identify them as:
\begin{equation}\label{loop of R}
    \mathcal {R}_{s,d-1}=\Omega(\mathcal {R}_{s,d}).
\end{equation}
Now it is straightforward to obtain the loop sequence $\mathcal {R}_{(s,d-i)}=\Omega^i(\mathcal {R}_{(s,d)})$, where $\Omega^i$ is the $i$-fold loop space operator.
Based on this property, the classifying spaces in any spatial dimensions with the same symmetry can be associated to the $d=0$ as follows:
\begin{equation}\label{loop space of zero}
    \mathcal {R}_{(s,d)}=\Omega^{-d}(\mathcal {R}_{(s,0})=\Omega^{-d}(\mathcal {R}_s)=\Sigma^d(\mathcal {R}_s).
\end{equation}
 \begin{table}
\centering
\caption{\textbf{Table of the classifying space $\mathcal {C}_s,\mathcal {R}_s$. The first column denotes the symmetric spaces.  The last column is the corresponding zero homotopy group, which displays the set parameterizing the disconnected regions of $\mathcal {C}_s$ and $\mathcal {R}_s$.}}\label{table2}
\begin{tabular}{ccc}
  \hline\hline
  s&Classifying Space  & $\pi_0$ \\\hline

       0&${U(2n)/(U(n)\times U(n))}\times \mathbb{Z}$& $\mathbb{Z}$\\
       1&$U(n)$ &0\\\hline
        0&$ \{O(2n)/(O(n)\times O(n))\}\times \mathbb{Z}$  & $\mathbb{Z}$\\
       1&$O(16n) $ &$\mathbb{Z}_2$\\
       2&$O(16n)/U(8n) $&$\mathbb{Z}_2$ \\
       3&$U(8n)/Sp(4n) $ &0\\
       4&$\{Sp(4n)/(Sp(2n)\times Sp(2n))\}\times\mathbb{Z} $& $\mathbb{Z}$\\
       5&$Sp(2n) $& 0\\
       6&$Sp(2n)/U(2n)  $&0\\
       7&$U(2n)/O(2n) $&0\\\hline\hline
\end{tabular}
\end{table}
\section{Loop sequence of $\mathcal {R}_{s,0}=\mathcal {R}_{s}$ between different symmetric classes}
  In zero dimension, the Hamiltonian is not dependent on the momentum $\textbf{k}$. So it is required to search for all possible symmetric spaces without considering the complexity affected by the involution of $\textbf{k}\rightarrow-\textbf{k}$ for $d\neq0$ case. The classifying spaces in zero dimension $\mathcal {R}_{s}$ have been studied by Bott\cite{bott1959stable} and Milnor\cite{milnor1963morse} in mathematics community.  Recently, M. Stone $et al$, from a more physical point of view, got the real classifying spaces by introducing more anti-commuting orthogonal complex structures (\ref{complex structure}) for the group $O(N)$, where $N$ is so large that the homotopy group $\pi_n(O(N))$ is in the stable range\cite{hou1997differential}. Physically, the classifying spaces can be understood as the manifold of the Goldstone mode after symmetry breaking. In order to reveal the loop sequences between them, $\Omega_{k}(O(N))$ is  defined as the space of complex structure $J$ set, which is anti-commuting with the fixed $\{J_i\}_1^{k-1}$ when denoting $\Omega_0(O(N))= O(N)$. Here, $\{J_i: i=1,\cdots ,k-1\}$ is the anti-commuting orthogonal complex structures which satisfy
\begin{equation}\label{complex structure}
    J_iJ_j+J_jJ_i=-2\delta_{ij}\mathbb{I}.
\end{equation}
From the definition of $\Omega_{k}(O(N))$, we have the following sequence of spaces
$$\Omega_{k}(O(N))\subset\Omega_{k-1}(O(N))\subset\cdots\subset\Omega_1(O(N))\subset\Omega_0(O(N)).$$
  After eight steps, the sequence of spaces repeats itself as $\Omega_{k+8}(O)=\Omega_k(O)$, called Bott periodicity of orthogonal group. The explicit homogenous manifold has been listed in Table\ref{table2}. It is known from the work by $\acute{\textrm{E}}$lie Cartan, there are only ten possible symmetric compact groups that can be served as the Hamiltonians of gapped free-fermion systems with  $\mathcal {R}_s=\Omega_{s-1}(O(N))$ \cite{milnor1963morse,stone2011symmetries} for $real$ case.
\par
Next, we will show the loop sequence of classifying spaces $\mathcal {R}_s$ by using minima geodesic construction method developed by Milnor\cite{milnor1963morse}. For the manifold of $\Omega_1$, one can derive that the $O(N)$ subspace satisfying $J_1^2=-1$ is homogenous manifold $O(2N)/U(N)$. In order to verify $\Omega(\Omega_0)=\Omega_1$, we construct a curve $L(\lambda)= e^{\lambda J_1}$. Because of $J_1^T= J_1^{-1}=-J_1$, one can get $e^{\lambda J_1^T}e^{\lambda J_1}=e^{\lambda (J_1^T+J_1)}=1,~ L(0)=\mathbb{I},~ L(\pi)=-\mathbb{I}$. Therefore, $L(\lambda)$ is a curve in the manifold $\Omega_0$ from the point $\mathbb{I}$ to $-\mathbb{I}$. Meanwhile, this curve is actually a minimal geodesic with its midpoint $L(\pi/2)= J_1$. Hence, the geodesic space can be represented by the space of $J_1$ satisfying $J_1^2=-1$. Since the space $\Omega_k(O)$ is symmetric, there indeed exists a loop curve $L(\lambda=0)=L(\lambda=2\pi)$ parameterized by $J_1$. Therefore we obtain $\Omega_1=\Omega(\Omega_0)$, or $\mathcal {R}_2=\Omega(\mathcal{R}_1)$.
\par
For the space $\Omega_2$ satisfying $J_2^2=-1$ and anticommuting with the fixed $J_1$, one can also get $\Omega_2 =\Omega(\Omega_1)$. Here we construct a curve $L(\lambda)= J_1e^{\lambda J_1^{-1}J_2}=J_1e^{\lambda A_1}$, where $J_2=J_1A_1$.  One will get $(J_1e^{\lambda A_1})^2 = J_1J_1J_1^{-1}e^{\lambda A_1}J_1e^{\lambda A_1}=-e^{\lambda(J_1^{-1} A_1J_1+A_1)}=-1$, $L(\lambda=0)=J_1$ and $L(\pi)=-J_1$. It is thus deduced that $L(\lambda)=J_1e^{\lambda A_1}$ is the minimal geodesic of $\Omega_1$ with the midpoint $L(\pi/2)=J_1A_1=J_2$. Obviously, the loop space of $\Omega_1$ is nothing but $\Omega_2$, or $\mathcal {R}_3=\Omega(\mathcal{R}_2)$.
\par
In general, one can prove the space $\Omega_{i+1}=\Omega(\Omega_i)$ by similar construction method described above. For example, we can define the curve as $L(\lambda)=J_ie^{\lambda J_i^{-1}J_{i+1}}=J_ie^{\lambda A_i}$, where $A_i=J_i^{-1}J_{i+1}$. It is easily to deduce that $A_i^2=-1$, and it anticommutes with $J_i$ , but commutes with $J_1,\ldots,J_{i-1}$.
So we have $L(\lambda)^2=-1$, with $L(\lambda)$ a submanifold of $\Omega_i$. Thus $L(\lambda)$ is a geodesic curve in $\Omega_i$ that interpolates between $L(0)=J_i$ and $L(\pi)=-J_i$, with $L(\pi/2)=J_{i+1}$. The set of geodesics can be parameterized by $\Omega_{i+1}$. And finally, we have $\Omega_{i+1}=\Omega(\Omega_i)$, or $\mathcal {R}_{i+2}=\Omega(\mathcal {R}_{i+1})$. By applying the Bott 8-periodic theorem on group $O(N)$, we obtain the following result:
\begin{equation}\label{Rs}
    \mathcal {R}_{s+1}=\Omega(\mathcal {R}_s), ~s~\textrm{mod}~ 8.
\end{equation}

\par
Though the construction method established above is tedious, its basic idea is relatively easy to follow: the loop space of $\Omega_i$ is approximated to the minimal geodesics (each loop can be homotopically deformed into the geodesic), and then these geodesics capture the topology of the $\Omega_{i+1}$ space. For example, the set of geodesics from the north to the south pole of the sphere is parameterized by the points on the equator of the sphere.

\section{Loop sequence of $\mathcal {R}_{s,d}$ for any symmetric classes and dimensions}
The loop sequence of any elements in Table\ref{table1} can be calculated by combining the equations (\ref{loop of R}) and (\ref{Rs}).  For example, we get $\mathcal {R}_{s+1,d}=\Omega(\mathcal {R}_{s,d})$ according to
\begin{equation}\label{rsd}
    \mathcal {R}_{s+1,d}=\Sigma^d(\mathcal {R}_{s+1})=\Sigma^d\Omega(\mathcal {R}_s)=\Omega(\mathcal {R}_{s,d}),
\end{equation}
 which means that the symmetric classifying spaces in any spatial dimension $d$ hold the same loop sequence structure as in the $d=0$ case discussed above.
And another very important result called (1,1) periodicity can also be derived as follows:
\begin{equation}\label{1,1 periodicity}
    \mathcal {R}_{s+1,d+1}=\Omega\Sigma(\mathcal {R}_{s,d})=\mathcal {R}_{s,d}.
\end{equation}
This is the key result of this work since it reveals the fundamental relationship within the (1,1) periodicity theorem of the real K-theory or KR-theory\cite{kitaev2009periodic}. After having the topological loop relationship for any classifying spaces, the gapped topological states of matters can therefore be classified by using homotopy group as $\pi_0(\mathcal {R}_{s-d})$, $s$ mod 8. It is obvious that increasing spatial dimension has the opposite effect on adding symmetry. So the distinct topological insulators in dimension $d$ is classified by $\pi_0(\mathcal {R}_{s-d})$, rather than $\pi_{d}(\mathcal {R}_{s})=\pi_0(\mathcal {R}_{s+d})$. In other words, it takes us back to the Bott clock\cite{stone2011symmetries}.
\section{conclusion}
In summary, the topological relationship between any classifying spaces for gapped free-fermion systems has been established with the loop sequence. And meanwhile, a simple homotopy explanation of the $d\rightarrow -d$ sign change affected by the inversion of $\textbf{k}\rightarrow-\textbf{k}$ of Bloch momentum has been given, which is usually very difficult to understand without using abstract K-theory. Even though the explicit Hamiltonian of the non-trivial topological states of matter is very complicated, our results show that the whole periodic table of topological classification has a relatively simple homotopy interpretation based on the loop sequence as developed in this paper.
\\\textbf{Acknowledgments}
\par
The author acknowledges the manuscript's reading and suggestions from Xinhui Zhang and Zhong Wang. This work is financially supported by the National Basic Research Program of China (No. 2011CB922200), and the National Natural Science Foundation of China (No. 11274302).
\\\textbf{Reference}
%\bibliography{topology}
%

\end{document}